\documentclass[manuscript,screen]{acmart}
\AtBeginDocument{%
  }

\usepackage{amsmath}  % 数学符号
\usepackage{stmaryrd}          % 提供一些额外的逻辑符号（可选）

\newcommand{\rotateRight}{\mathbin{\text{ROR}}}
\newcommand{\shr}{\mathbin{\text{SHR}}}

\setcopyright{acmlicensed}
\copyrightyear{2025}
\acmYear{2025}
\acmDOI{XXXXXXX.XXXXXXX}
\acmConference[Conference acronym 'XX]{Make sure to enter the correct
  conference title from your rights confirmation email}{June 03--05,
  2018}{Woodstock, NY}
\acmISBN{978-1-4503-XXXX-X/2018/06}

\begin{document}

\title{System Password Security: Attack and Defense Mechanisms}

\author{Chaofang Shi}
\affiliation{
  \institution{Hainan University}
  \country{China}
}
\email{wyscf163@163.com}
\authornote{Both Chaofang Shi and Zhongwen Li are co-first authors.}
\author{Zhongwen Li}
\authornotemark[1]
\affiliation{%
  \institution{Hainan University}
  \country{China}
}
 \email{lzw123@hainanu.edu.cn}

\author{Xiaoqi Li}
\affiliation{
  \institution{ Hainan University}
  \country{China}
}
\email{csxqli@ieee.org}


\begin{abstract}
 System passwords serve as critical credentials for user authentication and access control when logging into operating systems or applications. Upon entering a valid password, users pass verification to access system resources and execute corresponding operations. In recent years, frequent password cracking attacks targeting system passwords have posed a severe threat to information system security. To address this challenge, in-depth research into password cracking attack methods and defensive technologies holds significant importance. This paper conducts systematic research on system password security, focusing on analyzing typical password cracking methods such as brute force attacks, dictionary attacks, and rainbow table attacks, while evaluating the effectiveness of existing defensive measures. The experimental section utilizes common cryptanalysis tools, such as John the Ripper and Hashcat, to simulate brute force and dictionary attacks. Five test datasets, each generated using Message Digest Algorithm 5 (MD5), Secure Hash Algorithm 256-bit (SHA 256), and bcrypt hash functions, are analyzed. By comparing the overall performance of different hash algorithms and password complexity strategies against these attacks, the effectiveness of defensive measures such as salting and slow hashing algorithms is validated. Building upon this foundation, this paper further evaluates widely adopted defense mechanisms, including account lockout policies, multi-factor authentication, and risk adaptive authentication. By integrating experimental data with recent research findings, it analyzes the strengths and limitations of each approach while proposing feasible improvement recommendations and optimization strategies.
\end{abstract}



\maketitle

\section{Introduction}
\

In the era of the digital economy, information systems are deeply integrated into social production and daily life. As the first line of defence in access control, password authentication remains the most common authentication method due to its simple implementation and low cost\cite{byun2024security}. However, with the continuous evolution of attack techniques, password security faces unprecedented challenges. The computational power of high-performance brute-force cracking tools increases rapidly, large-scale dictionary libraries expand constantly, and the success rate of social-engineering attacks combined with machine learning rises significantly, so the traditional defence system undergoes a severe test~\cite{sen2012applied}. Therefore, the study of system password cracking attacks and defence techniques is of great theoretical and practical significance.
This research not only enriches the network security system at the academic level, but also provides direct guidance for practice\cite{touil2024efficient}. The systematic exploration of password attack mechanisms and defence mechanisms helps design a more secure password strategy and storage scheme, and enhances the public's security awareness and risk perception, thus promoting the construction of an overall network security culture\cite{he2024special}. At the same time, effective password defence technology applies to all kinds of information systems to reduce the risk of data leakage and business interruption, protect personal privacy and corporate assets, and strengthen the network security protection capability at the national level~\cite{bybee1986science}.
Based on the above background, this paper focuses on the typical attack methods of system password cracking and their defense strategies\cite{saadi2024survey}. Firstly, we review the research progress in the field of password security at home and abroad, and analyze the common methods and their technical characteristics, such as brute-force cracking, dictionary attacks, and rainbow table attacks, etc.~\cite{gong2025information}. Then, we design experiments, use John the Ripper and Hashcat tools to test the hash value of passwords generated by different hash algorithms (MD5, SHA-256, bcrypt) with brute-force and dictionary cracking, and evaluate the effectiveness of password complexity policy, salt, and slow hash algorithms\cite{he2024special}. Finally, we discuss new defense schemes, such as multi-factor authentication, risk-adaptive authentication, and honeywords, along with their optimization ideas in light of the latest research results~\cite{dell2010password}.
\section{Background}
\subsection{MD5}
\

\begin{figure}[ht]
    \centering    \includegraphics[width=0.8\linewidth]{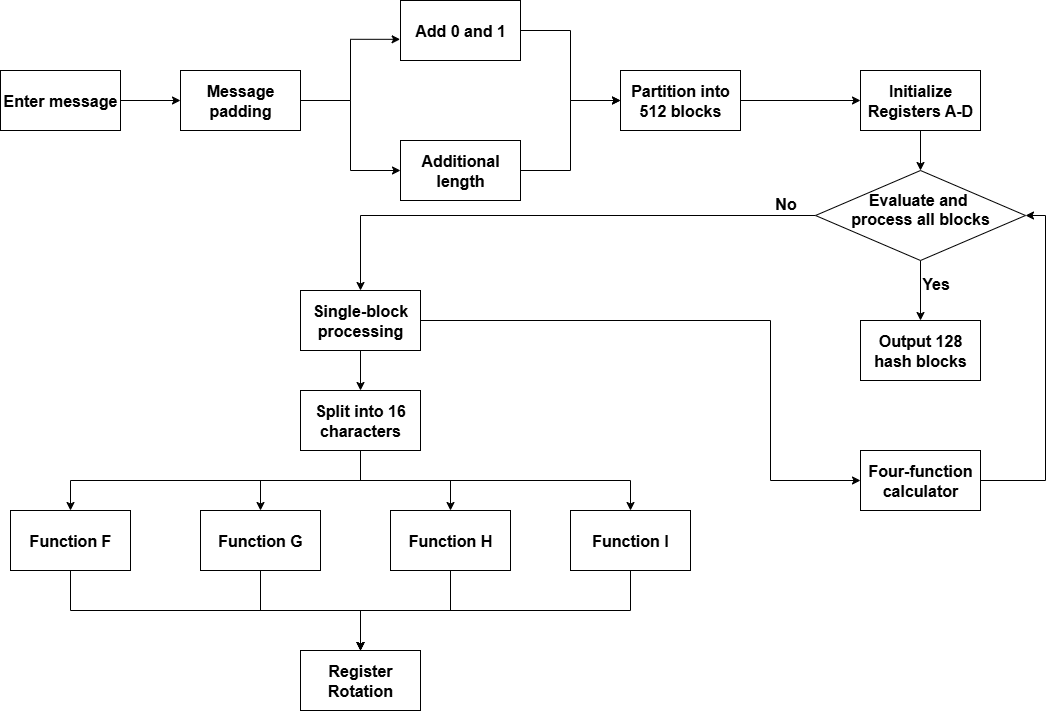}
    \caption{MD5 Algorithm Flowchart}
    \label{fig:1}
\end{figure}

The overall process of the MD5 algorithm is illustrated in Fig.~\ref{fig:1}. MD5 (Message-Digest Algorithm 5) is a widely used hash function that produces a 128-bit output\cite{alzahrani2024developing}. 

The basic procedure is as follows. The input message is first padded by appending a single "1" bit and enough "0" bits so that the total length is congruent to 448 modulo 512, and then a 64-bit little-endian representation of the original message length is appended\cite{almasoud2024chaotic}. After padding, the message is divided into multiple 512-bit blocks, each treated as 16 little-endian 32-bit words~\cite{bonneau2010password}.
The algorithm initializes four 32-bit registers $A$, $B$, $C$, and $D$ with the values
\[
A=0x67452301,\quad B=0xEFCDAB89,\quad C=0x98BADCFE,\quad D=0x10325476.
\]
For each data block, MD5 performs four rounds of nonlinear iterative operations with a total of 64 steps. Each round uses a distinct Boolean function $F$, $G$, $H$, or $I$, and a constant $T[i]$ derived from the sine function. Register updates follow the predefined order and left-rotation parameter $s$~\cite{lu2025movescanner}.
\[
A = B + \bigl((A + F(B,C,D) + M[k] + T[i]) \lll s \bigr),
\]
where $(A,B,C,D) \rightarrow (D,A,B,C)$ after each step.
After all data blocks are processed, the final values of $A$, $B$, $C$, and $D$ are concatenated in little-endian order to produce the 128-bit hash value\cite{zaikin2024inverting}. Although MD5 is computationally efficient, it is proven vulnerable to collision attacks and is no longer recommended for password storage or other security-critical applications. More secure hash algorithms, such as SHA-256 and bcrypt, are now preferred~\cite{bowen2009baiting}.

\subsection{SHA-256}
\

\begin{figure}[H]
    \centering
    \includegraphics[width=0.28\linewidth]{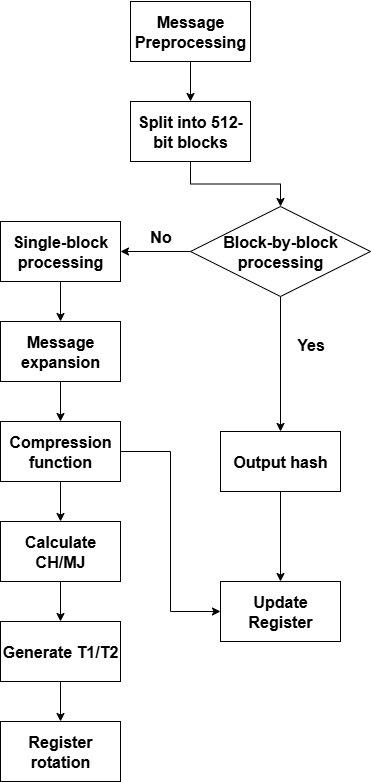}
    \caption{SHA-256 Algorithm Flowchart}
    \label{fig:2}
\end{figure}

SHA-256, as a cryptographic secure hash algorithm, belongs to the SHA-2 family.  
During the hashing process, the input data is first padded, with the original message length appended in big-endian order\cite{alimoglu2025edgchain}.  
The padded data is divided into 512-bit blocks, and each block is processed through 64 rounds of a compression function.  
In each round, different logical functions such as bitwise operations and modular addition, as well as predefined constants, are used to iteratively update eight 32-bit intermediate hash values~\cite{kong2025uechecker}.  
Finally, these eight intermediate values are concatenated to form a 256-bit hash.  
The core security of SHA-256 relies on its complex nonlinear structure and collision resistance\cite{li2024high}.  
Although no effective practical collision attack has yet been found, potential threats from quantum computing must still be considered~\cite{cohen2006use}. The SHA-256 algorithm process is shown in Fig.~\ref{fig:2}.

The implementation steps are as follows.

\textbf{(1) Message Preprocessing:} An input message $M$ of arbitrary length is padded and divided into $N$ blocks of 512 bits ($M_0$ to $M_{N-1}$).  
Padding requires the message length modulo $512$ to equal $448$, so a single bit "1" is appended, followed by enough "0" bits until the length reaches $448 \pmod{512}$.  
Finally, an $8$-byte big-endian representation of the original message length is appended\cite{wang2025ai}.

\textbf{(2) Hash Initialization:} Eight 32-bit registers $A$–$H$ are initialized as
$A = 0\text{x}6a09e667,\quad B = 0\text{x}bb67ae85,\quad C = 0\text{x}3c6ef372,\quad D = 0\text{x}a54ff53a,
E = 0\text{x}510e527f,\quad F = 0\text{x}9b05688c,\quad G = 0\text{x}1f83d9ab,\quad H = 0\text{x}5be0cd19.
$
These constants are derived from the fractional parts of the square roots of the first eight prime numbers~\cite{wang2019birthday}.

\textbf{(3) Message Expansion:} Each 512-bit block is split into 16 words $W_0, W_1, \dots, W_{15}$ and then expanded to 64 words $W_t$ with
\[
W_t = \sigma_1(W_{t-2}) + W_{t-7} + \sigma_0(W_{t-15}) + W_{t-16},\qquad 16 \le t < 64
\]
with
\[
\sigma_0(X) = (X \rotateRight 7) \oplus (X \rotateRight 18) \oplus (X \shr 3),
\]
\[
\sigma_1(X) = (X \rotateRight 17) \oplus (X \rotateRight 19) \oplus (X \shr 10),
\]
where $\rotateRight$ denotes cyclic right rotation and $\shr$ denotes logical right shift\cite{salih2024enhancing}.

\textbf{(4) Compression Function:} For each of the 64 rounds, the following Boolean functions are applied\cite{aydin2025novel}
\[
\Sigma_0(X) = (X \rotateRight 2) \oplus (X \rotateRight 13) \oplus (X \rotateRight 22),
\]
\[
\Sigma_1(X) = (X \rotateRight 6) \oplus (X \rotateRight 11) \oplus (X \rotateRight 25),
\]
\[
CH(X,Y,Z) = (X \land Y) \oplus (\lnot X \land Z),
\]
\[
MJ(X,Y,Z) = (X \land Y) \oplus (Y \land Z) \oplus (Z \land X).
\]
Two temporary variables are computed as

$
T1_t = H_t + \Sigma_1(E_t) + CH(E_t,F_t,G_t) + K_t + W_t,
T2_t = \Sigma_0(A_t) + MJ(A_t,B_t,C_t),
$

with $K_t$ denoting the $t$-th constant derived from the cube roots of the first 64 prime numbers\cite{zhou2025quantum}.  
The registers are updated according to
$
H_{t+1} = G_t,   
G_{t+1} = F_t,  
F_{t+1} = E_t, 
E_{t+1} = D_t + T1_t,
D_{t+1} = C_t, 
C_{t+1} = B_t, 
B_{t+1} = A_t, 
A_{t+1} = T1_t + T2_t.
$

\textbf{(5) Output:} After all message blocks are processed, the final 256-bit hash value is obtained by concatenating the final register values
\[
\mathrm{Hash}(M) = A_{\text{final}} \mathbin\| B_{\text{final}} \mathbin\| C_{\text{final}} \mathbin\| D_{\text{final}} \mathbin\| E_{\text{final}} \mathbin\| F_{\text{final}} \mathbin\| G_{\text{final}} \mathbin\| H_{\text{final}}.
\]

\subsection{Bcrypt algorithm}
\

\begin{figure}[ht]
    \centering
    \includegraphics[width=0.7\linewidth]{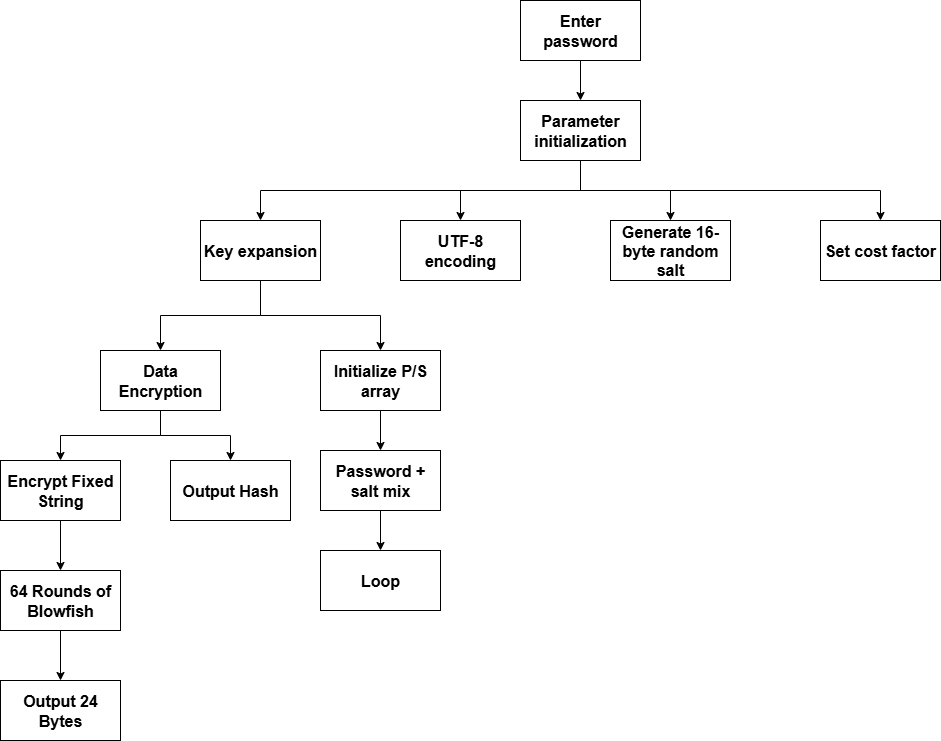}
    \caption{Bcrypt Algorithm Flowchart}
    \label{fig:3}
\end{figure}

Bcrypt is an adaptive hash algorithm specifically designed for password storage. Its core idea is to enhance resistance to brute-force and rainbow table attacks by introducing random salts, configurable computational cost, and memory-intensive operations\cite{al2025evaluasi}. The bcrypt algorithm process is illustrated in the figure. The final output is a 60-character string that includes the algorithm identifier, the cost value, a 22-character salt, and a 31-character hash. The security advantages of bcrypt can be summarized in three aspects~\cite{yegireddi2016survey}. First, the computational cost is adjustable because the cost parameter determines the number of iterations $(2^{\text{cost}})$, which enables the computation to scale with hardware performance. Second, a random salt is mandatory since a 16-byte (128-bit) cryptographically secure random salt is automatically generated for each encryption, which prevents rainbow table attacks\cite{guo2024cryptopyt}. Third, the computation is memory-intensive because the key expansion process requires substantial memory access, which helps resist GPU or ASIC acceleration~\cite{zhang2025penetration}. The bcrypt algorithm process is shown in Fig.~\ref{fig:3}.

\textbf{(1) Input and Initialization.} The user password is first encoded into a UTF-8 byte sequence $M$ and truncated to 72 bytes. 
The iteration parameter cost is set with a default value of 10, corresponding to $2^{10}$ iterations, and a 16-byte random salt is generated\cite{shi2025system}.

\textbf{(2) Key Expansion.}  
The P-array contains 18 subkeys of 32 bits each, initialized from the hexadecimal expansion of $\pi$. The S-box consists of four lookup tables of size $256 \times 32$ bits, also initialized from $\pi$.  
The password sequence $M$ is divided into 32-bit blocks $W_0 \ldots W_{m-1}$ in big-endian order, and cyclic padding is applied if necessary~\cite{peng2025multicfv}.  
The password and salt are alternately used to update the P-array and S-box through the EksBlowfishSetup algorithm
   \[
     P_i \leftarrow P_i \oplus W_{(i-1) \bmod m}
   \]
   Afterward, the Blowfish encryption function is repeatedly applied to mix in the salt $S$, and this process is repeated $2^\text{cost}$ times according to the cost parameter~\cite{jo2015new}.  

\textbf{(3) Data Encryption.} The fixed plaintext string "OrpheanBeholderScryDoubt" (24 bytes, three 64-bit blocks) is encrypted in CTR mode using 64 rounds of Blowfish\cite{castelo2024modification}. Each round consists of three steps. First, the P-array permutation is performed by XOR with the current subkey. Second, S-box substitution is carried out through non-linear table lookup and modular addition\cite{haryoputranto2025securing}. Third, the Feistel structure is updated according to
\[
L_r = L_{r-1} \oplus P_r,\quad 
F(L_r) = \bigl(S_1[b_0] + S_2[b_1]\bigr) \oplus \bigl(S_3[b_2] + S_4[b_3]\bigr) \bmod 2^{32},
\]
\[
R_r = R_{r-1} \oplus F(L_r),
\]
where $b_0, b_1, b_2, b_3$ are the four bytes of $L_r$.  

\textbf{(4) Output Format.} The three encrypted blocks are concatenated to produce a 24-byte ciphertext~\cite{zou2025malicious}. The final output string is represented as
\[
\$<\text{algorithm version}>\$<\text{cost value}>\$<\text{22-character salt}><\text{31-character hash}>
\]

\subsection{Password Cracking Attack}
\

In the field of information security, password cracking attacks evolve into multiple mature technical lines, each with distinct principles and application scenarios. With the advancement of computational power and the evolution of attack methods, modern password cracking exhibits characteristics of intelligence, automation, and scalability\cite{saputra2025password}. Attackers typically select the most effective single method according to the protection level of the target system or combine multiple methods to increase success probability~\cite{zhang2025risk}. Given the widespread deployment of online protections such as account lockout and human verification, attackers more commonly perform offline comparisons of stored password hashes to recover plaintext passwords. The following highlights several typical hash-targeted attack techniques~\cite{karuna2018generating}.

\textbf{(1) Brute-force attack} is the most primitive yet deterministic approach. Its core idea is to enumerate all possible character combinations to find a plaintext that matches the target hash. Although it is extremely time-consuming for high-complexity passwords, the method is universal and applicable to any password type\cite{reddyrobust}. Modern brute-force tools optimize search strategies through layered search that prioritizes high-frequency character combinations. They also leverage GPU parallelism or distributed computing to substantially increase search speed~\cite{shen2025blockchain}.

\textbf{(2) Dictionary attack} exploits statistical regularities and user habits. Attackers construct dictionaries containing common passwords, usernames, phrases, and their variants, such as case substitutions or numeric and symbol suffixes. These entries are compared against the target hash\cite{he2024special}. The effectiveness of the attack relies heavily on the quality of the dictionary and the rule engine. Advanced tools, such as those based on Markov chains or probabilistic models, infer user preferences from leaked data. This enables the cracking of many common passwords at a fraction of the computational cost of brute force in practice~\cite{temoshok2022digital}.

\textbf{(3) Rainbow table attack} applies a time--memory trade-off by precomputing and storing large mappings between hashes and candidate plaintexts using chain structures\cite{xu2025using}. Reduction functions map hash values back to candidate plaintexts, allowing attackers to trace a matching hash along a chain to potential original passwords. Compared to pure enumeration, rainbow tables reduce lookup time at the expense of storage overhead. Their effectiveness is significantly weakened when salts are incorporated into hash storage~\cite{wang2025ai}.

\subsection{Multi-Factor Authentication and Account Lock Mechanism}
\

Multi-factor authentication (MFA) requires users to provide two or more different types of verification factors when logging in. These factors are generally classified into three categories, including knowledge factors such as passwords, possession factors such as mobile devices and smart cards, and biometric factors such as fingerprints and facial recognition\cite{tihanyi2024privacy}. By combining independent elements, MFA adds a strong layer of defense so that an attacker who compromises one factor still cannot gain account access. Modern MFA implementations include time-based one-time passwords (TOTP), push-notification approvals, and biometric verification. Emerging passwordless technologies, such as the FIDO2 standard, integrate public-key cryptography with biometrics, improving both security and user experience. From a system-architecture perspective, advanced MFA solutions often adopt risk-based adaptive authentication, dynamically adjusting requirements according to contextual signals such as device fingerprints, geographic location, and behavioral patterns~\cite{omolara2018novel}.

The account lockout mechanism is an essential security measure designed to prevent unauthorized access and malicious attacks and represents one of the most effective countermeasures against brute-force attempts. Its principle is straightforward, as the system detects multiple consecutive incorrect password entries, abnormal login attempts, or other security risks and automatically locks the account, thereby blocking further login attempts. This mechanism makes it nearly impossible for attackers to compromise an account through brute-force or dictionary attacks, effectively safeguarding the confidentiality and integrity of user data and assets~\cite{liu2025empirical}. Account lockout is triggered automatically by the system or manually by an administrator. Once locked, a user must either wait for an automatic unlock after a specified period or perform additional identity verification to regain access. The process can also incorporate challenge-response mechanisms such as CAPTCHA to distinguish legitimate users from automated attack tools. Although highly effective, improper configuration of the lockout mechanism may be exploited by attackers to launch denial-of-service (DoS) attacks. To address this risk, modern account-protection systems detect distributed brute-force attacks and accurately identify and block malicious activity even when attackers use varying IP addresses and user agents~\cite{rao2006data}.
\vspace{-3ex}
\subsection{Salting Treatment and Risk-Adaptive Authentication}
\

Salted password hashing is a critical security technique for protecting stored credentials. Its core principle is to introduce a globally unique random salt for each user password, ensuring that even if different accounts share the same password, the resulting hash outputs are entirely distinct. Consequently, an attacker attempting a rainbow-table attack must construct a separate table for each user, fundamentally undermining the feasibility of such attacks. In practical deployment, combining salting with key derivation functions that resist brute force, such as PBKDF2, bcrypt, or Argon2, forces attackers to expend substantial time and computational resources even when equipped with high-performance hardware~\cite{schechter2009s}. The security of salted password hashing depends on three key requirements. First, the salt must be unpredictable and unique for each user. Second, algorithm cost parameters, including iteration count and memory requirement, must be chosen to increase the computational cost of attacks. Third, protective measures must be comprehensively integrated across storage, transmission, and verification. These measures jointly build a multi-layered defense against credential compromise. Numerous international standards bodies, including NIST, OWASP, and ISO/IEC, provide explicit guidance on salting, highlighting its critical role in password security~\cite{ross2005stronger}.

\begin{figure}[H]
    \centering
    \includegraphics[width=0.6\linewidth]{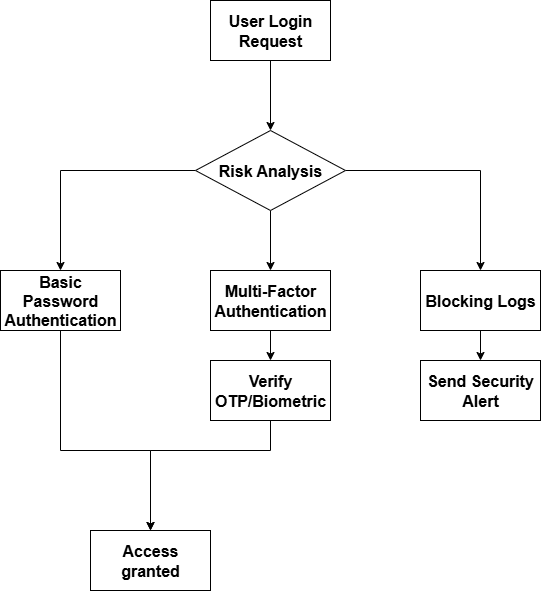}
    \caption{Risk-Adaptive Authentication System Workflow Diagram}
    \label{fig:4}
\end{figure}

Risk-adaptive authentication is an intelligent mechanism that dynamically adjusts authentication requirements based on real-time risk assessment. The system evaluates multiple contextual signals, including login time, geographic location, device fingerprint, and network environment, to generate a risk score for each login attempt. Low-risk requests require only password verification. Medium-risk requests require additional MFA. High-risk requests result in the login being blocked. During implementation, excessive blocking of legitimate users must be avoided. Providing an appeal or manual review channel is recommended to prevent unintended account lockouts~\cite{peng2025mining}. The overall operating principle of the risk-adaptive authentication system is illustrated in Fig.~\ref{fig:4}.

\section{Methods}

\subsection{Experimental Setup}
\

This work systematically evaluates the efficiency and success rates of dictionary and brute-force password attacks under varying conditions, providing empirical evidence for the design of password-defense mechanisms. The objectives are threefold. First, compare the performance and applicability of the two attack methods. Second, examine the cracking resistance of MD5, SHA-256, and bcrypt hash algorithms. Third, assess the practical effectiveness of password-complexity policies, salting, and slow hashing with bcrypt cost.

The experimental dataset comprises 50 passwords. They are partly drawn from public dictionaries to reflect real-world user habits and partly generated randomly to represent diverse levels of complexity. The password dataset selected for the experiment, along with its complexity analysis and selection rationale, is shown in the Tab~\ref{tab:1}. Experiments are conducted in two controlled environments. Environment (a) is a VMware virtual machine running VMware Workstation 17.5.1 build 23298084 with Kali Linux kernel 5.10.0-kali9-amd64 and John the Ripper 1.9.0-jumbo-1. Environment (b) is a Windows 11 system equipped with Intel(R) UHD Graphics and an NVIDIA GeForce RTX 3050 Laptop GPU running hashcat 6.2.6. The tools employ distinct traversal strategies. John uses predictive heuristics to characterize attack behavior. Hashcat enumerates from low to high keyspace to approximate real-world conditions. Passwords are hashed with MD5, SHA-256, and bcrypt. The bcrypt cost parameter is set to 10, which corresponds to 1024 rounds. Three dictionaries are used. Dictionary~1 is 801.4~KB, about 90,000 entries, the default Metasploit Kali common password list aggregated from community sources. Dictionary~2 is 133.4 MB, comprising approximately 10 million entries, and is derived from the 2009 RockYou data breach, containing many real-world weak passwords. Dictionary~3 is 14.6 GB and contains multi-source real-world passwords collected by security researchers. All experiments are performed under controlled conditions. Cracking time, success rate, and applied rules are recorded to ensure reproducibility and comparability of results.

\begin{table*}[htbp]
\centering
\caption{Selection Criteria for Test Datasets(Part of
the dataset)}
\resizebox{\textwidth}{!}{
\begin{tabular}{|c|c|c|p{9cm}|}
\hline
\textbf{Test Password} & \textbf{Complexity} & \textbf{Randomness} & \textbf{Selection Basis} \\ \hline
111111 & Low & Dictionary Password & Repeated digits, Top 10 password \\ \hline
112233 & Low & Dictionary Password & Sequential repeated digits \\ \hline
123456 & Low & Dictionary Password & Most common global password \\ \hline
123456789 & Low & Dictionary Password & Continuous numeric string \\ \hline
abcdef & Low & Dictionary Password & Sequential alphabetic string \\ \hline
vbjdsubv & Medium & Pseudo Random & Random-looking lowercase letters \\ \hline
@!! & Low & Pseudo Random & Single special character (non-standard) \\ \hline
!@\# & Low & Pseudo Random & Common special character combination \\ \hline
Qwerty123 & Medium & Dictionary Password & Keyboard pattern + numbers (weak mixed type) \\ \hline
10JQKA & Medium & Pseudo Random & Poker sequence (10, Jack, Queen, King, Ace) \\ \hline
AaBbCc & Medium & Pseudo Random & Alternating uppercase/lowercase letters \\ \hline
R14ERe & Medium & Pseudo Random & Letters + numbers combination \\ \hline
34fgdy & Medium & Pseudo Random & Digits + random lowercase letters \\ \hline
sfh3840 & Medium & Pseudo Random & Letters + digits combination \\ \hline
liziyu & Medium & Dictionary Password & Likely personal name \\ \hline
34vghd & Medium & Pseudo Random & Random digits and letters \\ \hline
JDVcnd & Medium & Pseudo Random & Random uppercase and lowercase letters \\ \hline
l56481 & Medium & Pseudo Random & Mixed letters and digits \\ \hline
fh6sdn & Medium & Pseudo Random & Random letters and digits \\ \hline
fsawdfs & Medium & Pseudo Random & Random lowercase letters \\ \hline
83740 & Low & Pseudo Random & Pure numeric string \\ \hline
56183152 & Low & Pseudo Random & Long numeric string \\ \hline
dhak1 & Medium & Pseudo Random & Letters + digits combination \\ \hline
Aa1!23 & High & Pseudo Random & Contains all types but short length (letters + digits + special chars) \\ \hline
P@ssw0rd! & High & Dictionary Password & Complex variant of “password” \\ \hline
M13rYnd4! & High & Pseudo Random & Mixed uppercase/lowercase + digits + special char \\ \hline
1w2e3r4t5y6u! & High & Dictionary Password & Keyboard pattern + digits + special char \\ \hline
SVJDNVKD & High & Pseudo Random & All uppercase random letters \\ \hline
\end{tabular}
}
\label{tab:1}
\end{table*}

\subsection{ Experimental Procedure}
\

The experimental procedure is as follows.

\begin{enumerate}
  \item \textbf{Dataset preparation.} A password dataset is collected, and each password is sequentially hashed using MD5 and SHA-256. The resulting hashes are categorized and stored in separate text files according to the encryption method (see Tab~\ref{tab:7}). Two independent tools are used in parallel to verify the correctness of every hash.
  
  \item \textbf{Initial cracking tests.} John the Ripper is employed to perform brute-force and dictionary attacks on password samples of varying strengths. For each hashing method, we record the cracking time, dictionary-attack success rate, intermediate guessing speed at different time points, and the total time required to crack each password group.
  
  \item \textbf{bcrypt hashing and verification.} Based on the results of step~(2), a subset of the dataset is encrypted using bcrypt with a cost parameter of~10 (see Appendix). After encryption, the bcrypt verification tool confirms the correctness of each hash, and the hashes are then subjected to cracking tests. Because bcrypt is computationally intensive, only the relatively easy subset from step~(2) is selected. Cracking time, real-time guessing speed, and total duration are recorded for each attack method.
  
  \item \textbf{Multithreaded evaluation.} To highlight bcrypt's resistance to parallel cracking, multithreading is introduced into all three attack methods, and the tests of step~(3) are repeated with full data logging.
  
  \item \textbf{High-difficulty bcrypt tests.} For bcrypt hashes that remain unbroken, six additional attack strategies, including multithreaded attacks, are performed individually, and John the Ripper's estimated cracking times are documented.
  
  \item \textbf{Hashcat brute-force.} Hashcat is applied to brute-force hashes produced by the three encryption methods. To better reflect real-world conditions, a mask mode with a 69-character set \(\{\texttt{a--z},\ \texttt{A--Z},\ \texttt{0--9},\ \texttt{!@\#\$\%\^\&*}\}\) is used, enumerating passwords from shorter to longer lengths while cracking times are recorded.
  
  \item \textbf{Data consolidation and analysis.} All experimental data are compiled, and key metrics are extracted for further analysis.
\end{enumerate}

\begin{table*}[htbp]
\centering
\caption{ Password Dataset Used in the SHA-256/MD5 Cracking Experiment and Their SHA-256/MD5-Encrypted Hash Values(Part of the dataset)}
\label{tab:password_hashes}
\resizebox{\textwidth}{!}{%
\begin{tabular}{|r|l|l|l|}
\hline
ID & Code & \texttt{SHA-256} & \texttt{MD5} \\
\hline
1  & 111111   & \texttt{\detokenize{BCB15F821479B4D5772BD0CA866C00AD5F926E3580720659CC80D39C9D09802A}} & \texttt{\detokenize{96E79218965E672C92A549DD5A330112}} \\
\hline
2  & 123123   & \texttt{\detokenize{96CAE35CE8A9B0244178BF28E4966C2CE1B8385723A96A6B838858CDD6CA0A1E}} & \texttt{\detokenize{4297f4461395523524562497399d7a93}} \\
\hline
3  & 123456   & \texttt{\detokenize{8D969EEF6ECAD3C29A3A629280E686CF0C3F5D5A86AFF3CA12020C923ADC6C92}} & \texttt{\detokenize{e10adc39496a59a66e56e057f20f883e}} \\
\hline
4  & 123456789 & \texttt{\detokenize{15E2B0D3C33891EBB0F1EF609EC419420C20E320CE94C65FBC8C3312448EB225}} & \texttt{\detokenize{25f9e794323b453885f5181f1b624d0b}} \\
\hline
5  & 200254   & \texttt{\detokenize{FA728F93B69CDB3E70B9EA482C0039524447C9A3D30E00A3B131500250A3F08B}} & \texttt{\detokenize{46a8f4ad1a350f1a44c94592e810d0c9}} \\
\hline
6  & 19780214 & \texttt{\detokenize{9D49A64404F625228E61DE6952FEEBA7CC21A41622AA8F8C473B1B9E67CFFCA6}} & \texttt{\detokenize{7f186571199334366639913ca6f1880d}} \\
\hline
7  & abcdef   & \texttt{\detokenize{BEF57EC7F53A6D40BEB640A780A639C83BC29AC8A9816F1FC6C5C6DCD93C4721}} & \texttt{\detokenize{e80b5017098950fc58aad83c8c14978e}} \\
\hline
8  & password & \texttt{\detokenize{5E884898DA2804715100E56F8DC6292773603D006AABBDD62A11EF72101542D8}} & \texttt{\detokenize{5f4dcc3b5aa765d61d8327deb882cf99}} \\
\hline
9  & iloveyou & \texttt{\detokenize{E4AD93CA07ACBBD908A3AA41E920EA4F4EF4F26E7F86CF8291C5DB28978045AE}} & \texttt{\detokenize{f25a2fc72690b780b2a14e140ef6a9e0}} \\
\hline
10 & dragon   & \texttt{\detokenize{A9C43BE948C5CABD56EF2BACFFB77CDAA5EEC49DD5EB0CC4129CF3EDA5F0E74C}} & \texttt{\detokenize{8621ffdbc5698829397d97767ac13db3}} \\
\hline
11 & sunshine & \texttt{\detokenize{A94144C4FD0C01CDDEF61B8BE963BF4C1E2B0811C037CE3F1835FDDF6EF6C223}} & \texttt{\detokenize{0571749e2ac330a7455809c6b0e7af90}} \\
\hline
12 & qwerty   & \texttt{\detokenize{65E84BE33532FB784C4812909EFF3A682B2716800EA744B2CF58EE0233705806}} & \texttt{\detokenize{d8578edf8458ce06fbc5bb76a58c5ca4}} \\
\hline
13 & admin    & \texttt{\detokenize{E5B5410415BDE908BD4D15DFB167A9C873FC4BBBA81F6F2AB448A918648AC76B}} & \texttt{\detokenize{21232f297a57a5a743894a0e4a801fc3}} \\
\hline
\end{tabular}%
}
\label{tab:7}
\end{table*}

\section{Evaluation}
\

This study systematically compares the efficiency of dictionary attacks and brute-force attacks under various conditions and achieves three main objectives. First, it confirms that brute-force cracking time grows exponentially with password strength, while the effectiveness of dictionary attacks is highly dependent on dictionary quality. Second, it demonstrates the insufficient security of fast hash algorithms such as MD5 and SHA-256. For example, eight-digit numeric passwords are cracked rapidly, whereas bcrypt, with its salt, adjustable cost parameter set to 10, and memory-intensive computation, significantly increases resistance to attacks. The brute-force speed for bcrypt is approximately one millionth that of MD5, and multithreading and GPU acceleration are largely ineffective, making bcrypt suitable for password storage. Third, it indicates that, in the absence of quantum computing, using bcrypt together with mixed passwords containing six or more characters provides strong security for password storage.

The experiment verifies the core principle of password security, which is to make the cost of an attack far exceed the value of the data, a goal that bcrypt achieves effectively. By comparing the efficiency of dictionary and brute-force attacks under different conditions and considering the effects of MD5, SHA-256, and bcrypt, the study also validates the effectiveness of defense measures such as password complexity policies, salting, and slow hashing, while highlighting the limitations of MD5 and SHA-256 for secure password storage.

Nevertheless, several limitations are noted. First, due to limited hardware, the experimental environment differs from that of a real attacker. The results mainly reflect the characteristics of the attack and defense methods rather than indicating that passwords not cracked in the experiment are secure in practice. Using higher-performance equipment better approximates real-world conditions and makes the results more convincing. Second, data variability must be considered. Although the John the Ripper experiments control variables through a virtual machine, cracking speeds fluctuate as the tool reaches its maximum performance, leading to minor differences in repeated measurements. Such variability could be amplified when cracking weaker passwords. Future research should adopt more rigorous datasets or counting methods to improve data accuracy and consistency.

\subsection{Hashcat Brute Force Experiment}
\
  
The experimental results are presented in the Tab~\ref{tab:10}. In this experiment, Hashcat is used to brute-force hashes generated from passwords of varying lengths and different hash algorithms. The results provide a realistic approximation of real-world attack scenarios and clearly illustrate the exponential increase in brute-force cost as password length grows. The data also highlight the significant security risks of using fast hash algorithms such as MD5 and SHA-256 for password storage. For passwords of equivalent strength, fast hashes require considerably less time to crack than slow hashes, making them more vulnerable. Although slow and salted algorithms such as bcrypt substantially increase cracking difficulty, insufficient password strength can still result in compromised credentials. These findings underscore that, in addition to employing appropriate slow hashing algorithms, ensuring adequate password length, complexity, and randomness is critical for effective protection.
\begin{table*}[htbp]
\centering
\caption{Time required for hashcat to brute-force hash values encrypted with different bit lengths and encryption methods}
\label{tab:10}
\resizebox{\textwidth}{!}{
\begin{tabular}{|c|c|c|c|c|}
\hline
\textbf{Password Length} & \textbf{Character Combination Count (unsalted)} & \textbf{Brute-force MD5 Time} & \textbf{Brute-force SHA-256 Time} & \textbf{Brute-force bcrypt Time} \\
\hline
1-digit password & 69 & $<$1s & $<$1s & 5min52s \\
\hline
2-digit password & 4,761 & $<$1s & $<$1s & 1h9min \\
\hline
3-digit password & $\approx 0.33$M & $<$1s & $<$1s & 4h27min \\
\hline
4-digit password & $\approx 22.7$M & $<$1s & $<$1s & 56h (estimated) \\
\hline
5-digit password & $\approx 15.6$B & $<$1s & 1min15s & 280 days (estimated) \\
\hline
6-digit password & $\approx 1079.2$B & 34s & 28min22s & 48 years (estimated) \\
\hline
7-digit password & $\approx 7.4$T & 28min22s & 1.7h & 48 years (estimated) \\
\hline
8-digit password & $\approx 51.3$T & 29h (estimated) & 117h (estimated) & -- \\
\hline
9-digit password & $\approx 3,540$T & 85h (estimated) & 1 year (estimated) & -- \\
\hline
\end{tabular}
}
\end{table*}

\subsection{Cracking MD5 Experiment}
\

This experiment validates the empirical model that defines password strength as the product of length, complexity, and randomness. The model is intended to emphasize the importance of these three factors without suggesting a linear relationship. The results show clear differences among attack methods. Brute force cracking time increases exponentially as password strength grows, while a large and well-designed dictionary can greatly accelerate cracking. However, the success of dictionary attacks still depends strongly on the quality of the dictionary. Storing all possible password combinations is neither practical nor meaningful because such an exhaustive search space would require unlimited storage and is essentially equivalent to brute force searching. In addition, since the MD5 algorithm is relatively simple, weak passwords provide almost no real protection. Even mixed character passwords of considerable length can be easily cracked when only protected by MD5, making this algorithm a major security risk in practice. The statistical results of MD5 cracking are presented in the Tab~\ref{tab:2}.
\begin{table*}[htbp]
\centering
\caption{Statistics on MD5 Cracking Experiment Results(part of the dataset)}
\vspace{-2ex}
\label{tab:2}
\resizebox{\textwidth}{!}{%
\begin{tabular}{|c|c|c|c|c|c|}
\hline
\textbf{No.} & \textbf{Password (MD5 processed)} & \textbf{Brute-force Time} & \textbf{Dictionary Attack 1} & \textbf{Dictionary Attack 2} & \textbf{Dictionary Attack 3} \\
\hline
1 & 111111 & $<$1s & 1 & 1 & 1 \\
\hline
2 & 123123 & $<$1s & 1 & 1 & 1 \\
\hline
3 & 123456 & $<$1s & 1 & 1 & 1 \\
\hline
4 & 123456789 & 0:00:23 & 1 & 1 & 1 \\
\hline
5 & 200254 & 0:00:01 & 1 & 1 & 1 \\
\hline
6 & 19780214 & 0:00:02 & 1 & 1 & 1 \\
\hline
7 & abcdef & $<$1s & 1 & 1 & 1 \\
\hline
8 & password & $<$1s & 1 & 1 & 1 \\
\hline
9 & iloveyou & 0:00:01 & 1 & 1 & 1 \\
\hline
10 & dragon & 0:00:01 & 1 & 1 & 1 \\
\hline
11 & sunshine & 0:00:01 & 1 & 1 & 1 \\
\hline
12 & qwerty & 0:02:51 & 1 & 1 & 1 \\
\hline
13 & admin & 0:00:01 & 1 & 1 & 1 \\
\hline
14 & vbjdsbu & 1:31:44 & 0 & 1 & 1 \\
\hline
15 & !! & 0:00:02 & 0 & 1 & 1 \\
\hline
16 & !@\# & 1:45:54 & 0 & 1 & 1 \\
\hline
17 & Qwerty123 & 0:08:57 & 0 & 1 & 1 \\
\hline
18 & abc123 & $>$8h & 1 & 1 & 1 \\
\hline
19 & password1 & 0:00:00 & 1 & 0 & 0 \\
\hline
20 & password123 & 2:05:27 & 1 & 0 & 0 \\
\hline
21 & 10JQKA & $>$8h & 0 & 0 & 0 \\
\hline
22 & AaBbCc & $>$8h & 0 & 0 & 0 \\
\hline
23 & R14ERe & 3:17:27 & 0 & 0 & 0 \\
\hline
24 & Aa!123 & $>$8h & 0 & 0 & 0 \\
\hline
25 & P@ssw0rd! & $>$8h & 0 & 0 & 0 \\
\hline
\textbf{Attack Completion Time} &  &  & 12h & $<$1s & 0:01:14 \\
\hline
\textbf{Real-time Cracking Speed (5min)} &  & 40960Kp/s &  &  &  \\
\hline
\textbf{Real-time Cracking Speed (10min)} &  & 41033Kp/s &  &  &  \\
\hline
\textbf{Real-time Cracking Speed (15min)} &  & 41150Kp/s &  &  &  \\
\hline
\textbf{Real-time Cracking Speed (20min)} &  & 41353Kp/s &  &  &  \\
\hline
\textbf{Real-time Cracking Speed (25min)} &  & 41440Kp/s &  &  &  \\
\hline
\end{tabular}%
} % end \resizebox
\end{table*}

\subsection{Cracking SHA-256 Experiment}
\

Because SHA-256 provides stronger collision resistance and a more complex hashing process than MD5, it is relatively harder to crack. This is reflected in the brute-force tests, where the password-guessing speed for SHA-256 is approximately 56 percent of that for MD5. However, this gap is not large enough to guarantee strong protection. In practical password storage, relying solely on SHA-256 hashing may still lead to credential compromise under sophisticated attacks. Even without GPU acceleration or multithreading, the experiment succeeds in cracking certain low-complexity passwords, for example, an eight-digit numeric password such as 12345678 and a seven-character lowercase and digit mixed password such as abc1234, within a few hours. In real-world scenarios, attackers typically possess more powerful hardware and employ more advanced algorithms and techniques, so using SHA-256 alone to protect passwords remains a significant security risk. The experimental results are shown in the Tab~\ref{tab:3}.

\begin{table*}[htbp]
\centering
\caption{John the Ripper Brute-Force SHA-256 Experiment Data(part of the dataset)}
\label{tab:3}
\resizebox{\textwidth}{!}{%
\begin{tabular}{|c|c|c|c|c|c|}
\hline
\textbf{No.} & \textbf{Password (SHA-256 processed)} & \textbf{Brute-force Time} & \textbf{Dictionary Attack 1} & \textbf{Dictionary Attack 2} & \textbf{Dictionary Attack 3} \\
\hline
1 & 111111 & \textless1s & 1 & 1 & 1 \\
2 & 123123 & \textless1s & 1 & 1 & 1 \\
3 & 123456 & \textless1s & 1 & 1 & 1 \\
4 & 123456789 & 0:00:23 & 1 & 1 & 1 \\
5 & 200254 & 0:00:01 & 1 & 1 & 1 \\
6 & 19780214 & 0:00:02 & 1 & 1 & 1 \\
7 & abcdef & 0:00:03 & 1 & 1 & 1 \\
8 & password & 0:00:01 & 1 & 1 & 1 \\
9 & iloveyou & 0:00:08 & 1 & 1 & 1 \\
10 & dragon & 0:03:08 & 1 & 1 & 1 \\
11 & sunshine & 0:00:01 & 1 & 1 & 1 \\
12 & qwerty &  & 1 & 1 & 1 \\
13 & admin &  & 1 & 1 & 1 \\
14 & vbjdsbu & 2:06:47 & 0 & 1 & 1 \\
15 & !! & 2:48:24 & 0 & 1 & 1 \\
16 & !@\# & 0:12:42 & 0 & 1 & 1 \\
17 & Qwerty123 & \textgreater12h & 1 & 1 & 1 \\
18 & abc123 & \textless1s & 1 & 1 & 1 \\
19 & password1 & 0:00:44 & 1 & 0 & 0 \\
20 & password123 & 3:30:13 & 1 & 0 & 0 \\
21 & 10JQKA & \textgreater12h & 0 & 0 & 0 \\
22 & AaBbCc & \textgreater12h & 0 & 0 & 0 \\
23 & R14ERe & 4:32:36 & 0 & 0 & 0 \\
24 & Aa!123 & \textgreater12h & 0 & 0 & 0 \\
25 & P@ssw0rd! & \textgreater12h & 0 & 0 & 0 \\
26 & M13YrNd4! & \textgreater\textgreater12h & 0 & 0 & 0 \\
27 & Q1we2r4t5y6u7i & \textgreater12h & 0 & 0 & 0 \\
28 & 34fgdg & 0:00:13 & 0 & 0 & 0 \\
\hline
\textbf{Attack Completion Time} &  & 12h & \textless1s & \textless1s & 0:01:22 \\
\hline
\textbf{Real-time Cracking Speed (5min)} &  & 25883Kp/s &  &  &  \\
\hline
\textbf{Real-time Cracking Speed (10min)} &  & 26178Kp/s &  &  &  \\
\hline
\textbf{Real-time Cracking Speed (15min)} &  & 26218Kp/s &  &  &  \\
\hline
\textbf{Real-time Cracking Speed (20min)} &  & 26259Kp/s &  &  &  \\
\hline
\textbf{Real-time Cracking Speed (25min)} &  & 26444Kp/s &  &  &  \\
\hline
\end{tabular}%
} % end \resizebox
\end{table*}

\subsection{Cracking bcrypt Experiment}
\

The password dataset and its encrypted hash values used in the bcrypt cracking experiment are shown in the Tab~\ref{tab:4}. The experimental results are also presented in the Tab~\ref{tab:9}.
In this experiment, the bcrypt algorithm is configured with a \textit{cost} parameter of 10. Approximate values indicated in the results are predicted by John the Ripper, based on the comparison between the password-guessing speed for individual hashes and the estimated average speed for cracking the entire dataset. 
The results demonstrate that bcrypt provides strong resistance against both brute-force and dictionary attacks due to its use of random salts. Its mechanism of frequent memory access rather than purely CPU-bound computation also nullifies the parallel advantage of multithreaded attacks. Additionally, the adjustable \textit{cost} parameter doubles the computational effort with each increment, causing brute-force cracking time to grow exponentially. These characteristics substantially increase the attack cost, making bcrypt a relatively secure choice for password storage.  
Nevertheless, weak passwords should still be avoided in practice. For example, simple passwords such as "123456," "dragon," and "123123" are cracked within seconds in our experiment. Comparisons of cracking time and password-guessing speed across different hashing algorithms are shown in the figure.  
Cracking bcrypt is far more difficult than cracking SHA-256 or MD5. In the experiment, the brute-force password-guessing speed for bcrypt is approximately one millionth that of MD5 and one of that of SHA-256. Consequently, some hashes that are cracked within seconds using MD5 or SHA-256 require nearly 10 hours when protected by bcrypt, even though their plaintext passwords are very simple. These results confirm that bcrypt offers strong security and is well-suited for password storage.

\begin{table*}[htbp]
\centering
\caption{Passwords and bcrypt Hash Values}
\label{tab:4}
\resizebox{\textwidth}{!}{%
\begin{tabular}{|c|c|c|}
\hline
\textbf{No.} & \textbf{Password} & \textbf{bcrypt Hash Value} \\
\hline
1 & 123456789 & \texttt{\$2a\$10\$Y0pz86lvVpYXDZgkrT/JheGQG8wTbPXfaRq2xBj.Ne16RxzZWXLfO} \\
\hline
2 & 111111 & \texttt{\$2a\$10\$9hosumvMZjK9eMVCI2qzeO0tWgyCYb5.2ULlCRkGFDSh1vk1r0jvu} \\
\hline
3 & 123123 & \texttt{\$2a\$10\$j7qZaTfmX87GrHHZfMKvludnjl2oYexEUL3SradS44BgZcz4opr4.} \\
\hline
4 & 123456 & \texttt{\$2a\$10\$cixOlw9mU.IT1QBxFZ1fduwdiZ7yUjIK0Wt5py9medOXnUKcleXjq} \\
\hline
5 & 200254 & \texttt{\$2a\$10\$I2CiytLTS0YFs7mCPdjQsexwRzykbjKVSNUGaP7MMUSiMfEeobkdG} \\
\hline
6 & 19780214 & \texttt{\$2a\$10\$y/yNWjKt2l/e00M/bsavNeHno9jzF39W721iR6ujMWfnaHcVtTV8a} \\
\hline
7 & abcdef & \texttt{\$2a\$10\$Okog3OBiZ664q73iYptT2eaKOAAzTRohPSkSMIEZYSkali7Sk/o5q} \\
\hline
8 & dhak1 & \texttt{\$2a\$10\$4LfH3264Jk2lu42S5LSKreseWCzwHdETMySHxJwEKJD06QeRtPhS6} \\
\hline
9 & dragon & \texttt{\$2a\$10\$4SCiziTjExT5lzdB.3yna.pmr1Hh4XK.a.xLentHl6TtlvNAZHQxO} \\
\hline
10 & !@\# & \texttt{\$2a\$10\$08BCj3miPJu9lVE8bRhb8e0XaNQIApE3oNeQ1TZAaA8EdlaXLzRba} \\
\hline
11 & Acj816 & \texttt{\$2a\$10\$IHVZXyzuOvtVeSHtYnLVy.H0aS4unReMCIVO3S9lgxNOcwqiQs17S} \\
\hline
12 & sfh3840 & \texttt{\$2a\$10\$3etSaX2jMzifNlwO0JNQeuHrjrelzfQV0O9hkx4.SuEyTyY96ot8Wk6} \\
\hline
13 & 34vgdh & \texttt{\$2a\$10\$.2gbfBTZRxzoWgrYggQBbO1DEn4UA/AzocILRZXg2mVlbqlcAR38O} \\
\hline
14 & iloveyou & \texttt{\$2a\$10\$kDYGRIV.WlArimXGHIUFDO7RQXp3Af3GLGovT3qdhdghWhOFomZVS} \\
\hline
15 & qwerty & \texttt{\$2a\$10\$.EblGu7kZg3JgXbxiiQZt.fsT7OXrTeJa1JrSJ2Bbp1XVVYJWfpXy} \\
\hline
16 & dhf576 & \texttt{\$2a\$10\$whwDXfJEyGEGXnt1aeFSHeGhtpphRctQeldJjGkx9dDQEuAuHN3za} \\
\hline
17 & 83740 & \texttt{\$2a\$10\$PS4nv60AAFffoecAjzMItOd0d4vCw5Qbwn.HJkFnC073Cs0evQuN2} \\
\hline
\end{tabular}%
} % end \resizebox
\end{table*}

\begin{table*}[htbp]
\centering
\caption{Experimental Results of bcrypt Hash Cracking and Dictionary Attacks}
\resizebox{\textwidth}{!}{
\begin{tabular}{|c|c|c|c|c|c|}
\hline
\textbf{No.} & \textbf{Password} & \textbf{bcrypt (processing time / brute-force cracking time)} & \textbf{Dictionary Attack} & \textbf{Dictionary Attack (4 threads)} & \textbf{Dictionary Attack 2 (4 threads accelerated)} \\
\hline
1 & l@\# & $\approx$9h & $\approx$9h & $\approx$10h & 0.28.26 \\
\hline
2 & Acj816 & $\approx$9h & $\approx$9h & $\approx$103h & 0.23.47 \\
\hline
3 & sfh3840 & $\approx$9h & $\approx$9h & $\approx$115h &  \\
\hline
4 & 34vghd & $\approx$9h & $\approx$18h &  &  \\
\hline
5 & iloveyou & $<1$s & 14s & $<1$s & $<1$s \\
\hline
6 & qwerty &  &  & 1s & 2s \\
\hline
7 & dhf576 & $\approx$9h & $\approx$15.8h &  &  \\
\hline
8 & 83740 & $\approx$9h & $\approx$9h &  &  \\
\hline
\textbf{Experimental Hashcat Speed (15min)} & 27.23p/s & $\approx$28.654p/s & 28.88p/s & $<28.602p/s$ & 20.38p/s \\
\hline
\textbf{Experimental Hashcat Speed (10min)} & 27.46p/s & $\approx$29.099p/s & 27.28p/s & $<29.241p/s$ & 20.48p/s \\
\hline
\textbf{Experimental Hashcat Speed (15min)} & 27.38p/s & $\approx$29.110p/s & 27.29p/s & $<29.547p/s$ & 20.42p/s \\
\hline
\textbf{Experimental Hashcat Speed (20min)} & 27.28p/s & $\approx$29.306p/s & 27.26p/s & $<29.659p/s$ & 20.43p/s \\
\hline
\textbf{Average Speed (Hashcat)} & $\approx$27.33p/s & $\approx$29.04p/s & $\approx$27.43p/s & $\approx$29.26p/s & $\approx$20.43p/s \\
\hline
\textbf{Estimated Cracking Time of Combined Dataset} & $\approx$96.2h & $\approx$95.5h & $\approx$117h & $\approx$151h & $\approx$27h \\
\hline
\end{tabular}
}
\label{tab:9}
\end{table*}

\section{Password Defense Strategy and Optimization}
\subsection{Password Complexity}
\

Password complexity policies serve as the first line of defense, aiming to ensure that user-chosen passwords possess sufficient strength to resist guessing or cracking. Typical requirements include minimum length, inclusion of uppercase and lowercase letters, digits, and special characters, avoidance of usernames, common words, or personally identifiable information, prohibition of password reuse, and encouragement of periodic updates. Advanced policy engines often utilize probabilistic assessment or machine learning methods to provide real-time strength scoring and visual feedback, guiding users to create passwords that are both secure and memorable~\cite{xiang2025security}.  

Experimental results from this study indicate that increasing password complexity—particularly length—substantially raises the cost for attackers. However, excessive complexity can reduce memorability, leading users to adopt unsafe coping strategies, such as reusing similar passwords after frequent resets or using the same password across multiple accounts, thereby weakening security. A practical alternative is the use of unrelated passphrases: combining multiple unassociated words or phrases into a longer yet memorable password increases the effective character space and reduces predictability, balancing security and usability~\cite{schechter2010popularity}.  

Based on the findings and current standards, it is recommended to follow the modern password guidelines from NIST SP 800-63B. These guidelines advocate abandoning rigid requirements for forced character mixing and complex patterns, and instead emphasize a minimum length (e.g., at least eight characters), banning common or previously compromised passwords, and providing real-time strength assessment and feedback. The guidelines note that traditional complexity rules may inadvertently increase predictability and encourage weak user behavior. Finally, user education and multi-factor authentication are implemented as complementary measures to further enhance overall authentication security without imposing a significant additional burden on users~\cite{spaord1992observations}.

\vspace{-4ex}
\subsection{ Honeywords}
\

Honeywords represent an active defense mechanism in which decoy passwords that are highly similar to the real passwords are injected into the password storage system to create a honeynet targeting potential attackers. When the database is compromised, attackers cannot distinguish real password hashes from decoy hashes. Attempts to use decoy passwords during offline cracking or login trigger an alert and allow system administrators to take countermeasures. For online attacks, attempts using Honeywords receive uniform error messages while being silently logged, which guides attackers into a honeypot environment for tracking and forensics~\cite{li2025interaction}.  

\begin{figure}[ht]
\vspace{-2ex}
    \centering
    \includegraphics[width=0.5\linewidth]{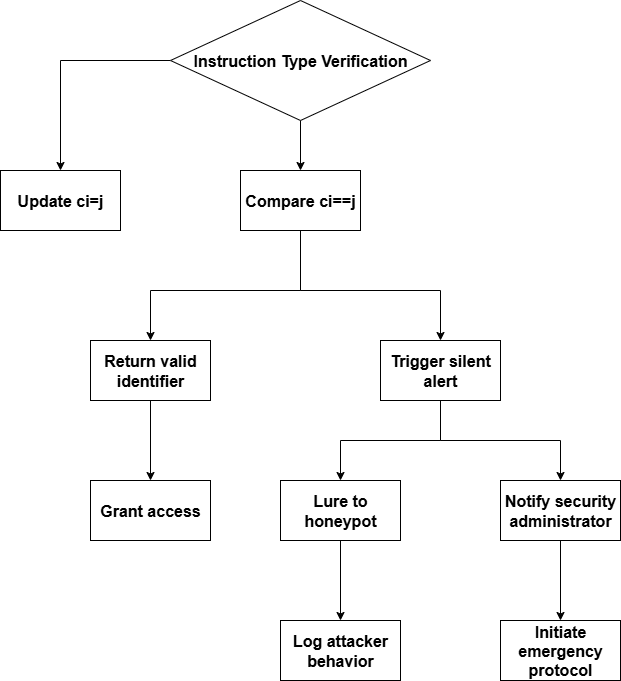}
    \caption{Honeywords Workflow Diagram}
    \label{fig:5}
\end{figure}

To enhance the credibility of decoy passwords and improve overall defense, Honeywords are often combined with slow hash algorithms such as bcrypt and use techniques such as keyboard-adjacent character substitution, GAN-based sample generation, or historical password transformation to dynamically produce realistic decoys. The position of the real password is protected through encrypted markers. Although this approach increases memory usage, the overhead remains acceptable in modern systems with high storage capacity and relatively low storage costs~\cite{spitzner2003honeytokens}. The working principle and workflow of Honeywords are illustrated in Figure~\ref{fig:5}.  

Honeywords are commonly used together with a Honeychecker to form a distributed defense mechanism. The main system stores a salted hash list that contains one real password and \(k-1\) decoy passwords, while the index of the real password is encrypted and stored in a physically isolated Honeychecker server. During authentication, the main system first performs basic verification, and when a match is found and the index corresponds to a decoy, it queries the Honeychecker, triggering a silent alert or redirecting the attacker to a honeypot~\cite{niu2025natlm}. Otherwise, a standard error message is returned. This design ensures that even if the primary database is compromised, attackers cannot identify the real credentials, and any use of decoy passwords directly reveals malicious intent. The architecture follows the distributed security principles described in NIST~SP~800-63B and adopts secret distribution and minimal verifier interfaces, which have been successfully applied in high-risk domains such as finance~\cite{menezes2018handbook}.

\subsection{Client-Side Key Derivation}
\

Client-Side Key Derivation (CSKD) is a technique in which cryptographic keys are generated and stored locally on the client, reducing server-side risks and improving the overall security of password storage and authentication. In practical deployments, a user-provided password is combined with a randomly generated salt and processed through multiple iterations of a one-way hash function using a key derivation function (KDF) such as PBKDF2, bcrypt, or Argon2~\cite{zhou2025blockchain}. These KDFs are intentionally designed to be computationally intensive, which greatly increases the time and resources required for brute-force attacks. 

The security of CSKD depends on the irreversibility of the hash functions and the randomness of the salt. Once a key is derived, the original password cannot be recovered from it, and the use of different salts ensures that identical passwords generate distinct derived keys for different users. Multiple iterations further increase computational cost and make brute-force attacks practically infeasible~\cite{rivest1992md5}.  

Despite its strong security advantages, CSKD introduces certain trade-offs. The derivation process adds computational overhead on the client, which may affect performance on resource-constrained devices such as IoT endpoints or low-end mobile devices. In addition, poor selection or misconfiguration of the KDF, such as using an insufficient iteration count or a weak algorithm, can create new security vulnerabilities.  

\begin{figure}[ht]
    \centering
    \includegraphics[width=0.4\linewidth]{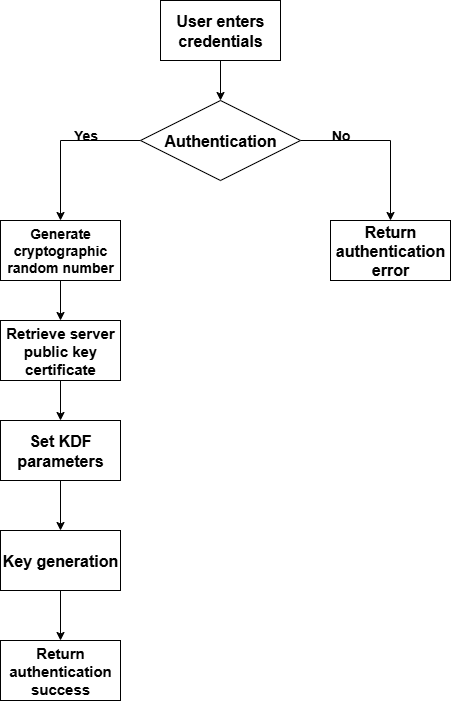}
    \caption{Client Key Derivation Schematic}
    \label{fig:6}
\end{figure}

Overall, CSKD strengthens password security by increasing key complexity and computational difficulty while reducing the need for users to memorize highly complex passwords~\cite{saarinen2003cryptanalysis}. In a typical implementation, the password and salt are combined and repeatedly hashed before being processed through a key derivation function to enhance protection. The underlying principles and workflow of CSKD are illustrated in Figure~\ref{fig:6}.

\subsection{Password Replacement Based on Biometrics}
 
Biometric-based authentication methods utilize unique physiological or behavioral characteristics such as fingerprints, facial features, iris patterns, or behavioral biometrics to replace or complement traditional passwords. This approach provides substantial improvements in both security and usability by eliminating the need for users to memorize complex passwords. When combined with liveness detection or dynamic challenge mechanisms, biometric authentication effectively resists brute-force attacks and phishing attempts~\cite{jin2025blockchain}. It operates either as an independent authentication mechanism or as a component of multi-factor authentication, as in smartphone fingerprint unlocking or facial recognition payments.  

Despite these advantages, biometric solutions present critical risks and operational constraints. Once biometric data are compromised, they cannot be modified, which necessitates encrypted storage and processing within trusted execution environments or secure hardware modules~\cite{yoshida2005analysis}. Compliance with privacy and regulatory frameworks is mandatory, including requirements for data minimization, purpose limitation, and informed user consent. Moreover, techniques such as liveness detection, replay attack prevention, and differential privacy are employed to mitigate the risks of forgery, side-channel attacks, and unauthorized use.  

Overall, biometric authentication enhances both user experience and system security, yet its deployment must be carefully designed with consideration for technical robustness, privacy protection, and regulatory compliance~\cite{narayanan2005fast}.

\section{Conclusion}
\

This study systematically reviews the principles and encryption processes of MD5, SHA-256, and bcrypt, and analyzes the characteristics of three common password attacks: brute-force, dictionary, and rainbow table attacks. The effectiveness of password defense measures, including password complexity policies, salting, and slow hash algorithms, is also evaluated. In addition, existing defenses such as multi-factor authentication, account lockout mechanisms, and risk-adaptive authentication are discussed. Emerging defenses, including Honeywords, Client-Side Key Derivation (CSKD), and biometric-based password replacement, are introduced, and the use of uncorrelated mixed passphrases is recommended to enhance individual password security.  
Experimental results indicate that traditional dictionary- and rainbow-table-based attacks remain efficient, particularly when modern attackers leverage social engineering, personal information analysis, and AI-based password inference. Patterned passwords still carry risks of being compromised. Moreover, password security often depends on overlooked vulnerabilities, and even sophisticated defenses cannot cover all weak points.

\bibliographystyle{unsrt}
\bibliography{reference}
\end{document}